# Variation of Contact Resonance Frequency during Domain Switching in PFM Measurements for Ferroelectric Materials

*Yue Liu, Yao Sun, Wanheng Lu, Hongli Wang, Zhongting Wang, Bingxue Yu, Tao Li, Kaiyang Zeng\**

Department of Mechanical Engineering, National University of Singapore, 9 Engineering Drive 1, Singapore 117576

ABSTRACT Piezoresponse Force Spectroscopy (PFS) is a powerful method widely used for measuring the nanoscale ferroelectric responses of the materials. However, it is found that certain non-ferroelectric materials can also generate similar responses from the PFS measurements due to many other factors, hence, it is believed that PFS alone is not sufficient to differentiate the ferroelectric and non-ferroelectric materials. On the other hands, this work shows that there are distinct differences in contact resonance frequency variation during the PFS measurements for ferroelectric and non-ferroelectric materials. Therefore, a new, simple and effective method is proposed to differentiate the responses from the ferroelectric and non-ferroelectric materials, this new analysis uses contact resonance frequency responses during the PFS measurements as a new parameter to differentiate the PFS measured responses from different materials.

Development and applications of the ferroelectric materials have been one of the most active topics for decades. Due to the unique characteristics of spontaneous polarization, ferroelectric materials have been used in a wide range of applications, such as sensors, actuators and memory devices.[1] Developing new ferroelectric materials has great significances for research and applications in the area of functional materials.[2] Comparing with the common dielectric materials with a linear polarization response, ferroelectric materials demonstrate a nonlinear and nonzero polarization response.[3] To study the ferroelectric phenomena at nanoscale, such as at domain level, Piezoresponse Force Microscopy (PFM) and its spectroscopy form, Piezoresponse Force Spectroscopy (PFS), are widely used in the last decades. As the premier characterization tools for domain structures, orientation and properties of the ferroelectric materials, PFM and PFS techniques can probe time- or voltage-dependent phenomena with high spatial resolution.[4] In the PFS measurements, the surface of the sample contacts with a sharp conductive tip at the end of PFM cantilever. After applying excitation of DC voltage and scanning of the sample surface with the same tip, local polarization switching may occur and can be detected by the same tip. However, due to the principle of probing method in PFS,[5,6] the measurements of the local ferroelectric responses can be affected by a number of factors. Besides the polarization-electric field (P-E) relationship, the electrostatic force between the tip and sample surface,[7] surface charging,[8–10] Vegard effect[11] and ionic mechanisms[12–14] can also induce the hysteresis-like loops in which are similar to the P-E loops obtained in ferroelectric materials during the PFS measurements. In addition, it is also noted that such hysteresis-like loop can also be observed in a broad variety of non-ferroelectric materials during the PFS measurements, for example, glass,[15] $LiCoO_2$,[12] $TiO_2$[16] and even banana peel.[5] It was therefore believed that the hysteresis loops obtained by PFS is insufficient as the only proof of the ferroelectricity.[17] Due to these facts, numbers of other methods to probe the local ferroelectric phenomena have been developed in recent years. These methods usually introduce different techniques other than Switching Spectroscopy Piezoresponse Force Microscopy (SS-PFM) and/or PFS to investigate the ferroelectric characteristics. For example, optical second harmonic generation (SHG) allows for differentiating ferroelectric and magnetic phase transitions by using the light beams with different incident wavelength.[18,19] Ultraviolet Raman Spectroscopy[20] and unit-cell scale mapping[21] also provide evidence for nanoscale ferroelectricity. On the other hand, contact Kelvin Probe Force Microscopy (cKPFM)[22] and frequency dependent PFM[23] are developed as the effective new measurements to differentiate the true ferroelectricity contributions with the combination of hysteresis loops in PFS measurements. Furthermore, various techniques with higher harmonic frequencies are also developed to distinguish the responses from the ferroelectric and non-ferroelectric materials.[11] Most of those experimental techniques are relatively complicated and require new set-ups, methods or analysis, because the PFM/PFS technique alone is insufficient to determine if the responses are real ferroelectric for an unknown material. On the other hand, almost all of the PFS or SS-PFM studies only analyze the amplitude and the phase angle changes induced by the external electric field, other parameters during the PFS measurements are largely ignored. Especially, the contact resonance frequency ($f_0$) and quality factor ($Q$) obtained during the PFS and SS-PFM measurements are not considered in the analysis published so far.

In this study, we first report the experimental observations of the changes of $f_0$ signals for a ferroelectric material, Pb(Zn$_{1/3}$Nb$_{2/3}$)O$_3$–9%PbTiO$_3$, or PZN-PT, and a non-ferroelectric material, banana peel, during the PFS measurements. It is found that there is a significant divergence of $f_0$ signal between the two materials. Therefore, a simple yet effective method has been proposed based on these observations, in which can differentiate ferroelectric material and non-ferroelectric material by means of the changes of the mechanical properties during the domain switching processes that are reflected by the variation of $f_0$ in the PFS measurements. This new method is then applied to overall four ferroelectric materials and four non-ferroelectric materials in future experiments to verify the new proposed analysis methodology.

RESULTS AND DISSCUSION

**Theory basis.** The contact resonance frequency of the tip and sample is mainly related to the mechanical properties of the cantilever and the tip-sample contact stiffness.[24–26] During the PFM measurements, the oscillation of cantilever is indirectly driven by the AC bias-induced sample surface oscillation.[27–29] Hence the instantaneous position of the tip in the vertical direction, $z$, obeys the driven damped harmonic oscillator equation as following:[30]

$$m_c \frac{dz^2}{d^2t} = -k_c z - c_c \frac{dz}{dt} + F_{st} + F_d \cos \omega_d t \tag{1}$$

where $F_d$ and $\omega_d$ are the amplitude and the angular frequency of the excitation force, respectively; $m_c$, $k_c$ and $c_c$ are the effective mass, the spring constant and the viscous damping coefficient of the free cantilever, respectively; and $F_{st}$ is the tip-sample interaction force and it is mostly attributed by Hertzian contact force. This force can be expressed as:[27]

$$F_{st}(z) = \frac{4}{3} E^* \sqrt{R} (a_0 - z - z_c)^{\frac{3}{2}} \tag{2}$$

with

$$\frac{1}{E^*} = \frac{(1-v_t^2)}{E_t} + \frac{(1-v_s^2)}{E_s} \tag{3}$$

where $R$ is the radius of the tip and $E^*$ is the effective Young's modulus of the tip-sample contact system, and this quantity is related to the Young's modulus of the tip ($E_t$) and sample ($E_s$), also the Poisson's ratio of the tip ($v_t$) and sample ($v_s$); and finally, $z_c$ is the equilibrium position of the cantilever. From Eqs. (2) and (3), it is obvious that the tip-sample interaction force is only related to the sample's mechanical properties when the tip is at a certain height. This driven harmonic oscillator can be simplified by a damping harmonic oscillator (DHO) model[24,27] which is driven by the amplitude ($A_d$) and phase ($\phi_d$) of the sample surface. In this case, the driving forces are transferred to a spring ($k^*$) and a dashpot ($c$) model in the system as showed in Figure 1. The spring constant, $k^*$, is related to $E^*$ by the following relation:[31]

$$k^* = 2E^* r_c \tag{4}$$

where $r_c$ is the radius of the contact area in Hertz indentation model. By monitoring the oscillation of the tip, the PFM measurements can obtain the oscillation parameters, including amplitude ($A$), phase angle ($\phi$), resonance frequency ($f_0$) and quality factor ($Q$). The $f_0$ is closely related to the ratio between the contact stiffness and the stiffness of the free cantilever ($k^*/k_c$). When $k^*/k_c$ increases, the normalized contact resonance frequency, $f_0/f_{0,c}$ (the ratio between the $f_0$ and the free cantilever frequency, $f_{0,c}$,) shifts from the free vibration to the clamped one, when the $k^*/k_c$ value is over 100, $f_0/f_{0,c}$ arises significantly.[27] The relation between $f_0$ and $k^*$ is:

$$\frac{f_0}{f_{0,c}} = \sqrt{\frac{k_c + k^*}{k_c}} \tag{5}$$

The viscous coefficient $c$ is mainly related to the output signals of $Q$ which indicates the dissipative energy.[24,32] The quality factor of a free PFM cantilever is caused by sound radiation and friction with air.[27] In many operation modes of the Scanning Probe Microscopy (SPM), $Q$ factor also reflects the mechanical and electric properties of the sample surface.[33,34]

In order to more accurately track $f_0$ during the PFM measurements, multifrequency techniques, including dual AC resonance tracking (DART) and band excitation (BE) are developed. In particular, DART technique modulates the tip-sample contact at two frequencies ($f_1$, $f_2$) where $f_0$ is located between the two. Each carrier frequency ($f_1$, $f_2$) has the corresponding amplitude ($A_1$, $A_2$) and phase ($\phi_1$, $\phi_2$). Hence, $f_0$ and $Q$ can be calculated from the measurements of $A_1$, $A_2$, $\phi_1$ and $\phi_2$.[24] In the PFM measurements, this resonance enhancement technique is developed to amplify the weak electromechanical response signals[35] through minimizing the interference between the contact resonance and the phase transition, and also enhancing the signal to noise ratio (SNR).[36] PFS measurement also tracks $f_0$ as an indicator of the mechanical properties, because the contact stiffness and the contact damping response are associated with the signals of $A$, $\phi$, $f_0$ and $Q$,[28] therefore, the mechanical transformation processes during the polarization switching in ferroelectric materials can be monitored during the PFS measurements. It is anticipated that, during the PFS measurements, mechanical properties during the domain switching processes should be different between the ferroelectric and non-ferroelectric materials due to the factors of domain switching in ferroelectric materials and the effects of other factors (non-ferroelectric domains) in the non-ferroelectric materials, which will be then revealed in the signal spectrum of the contact resonance frequency, $f_0$.

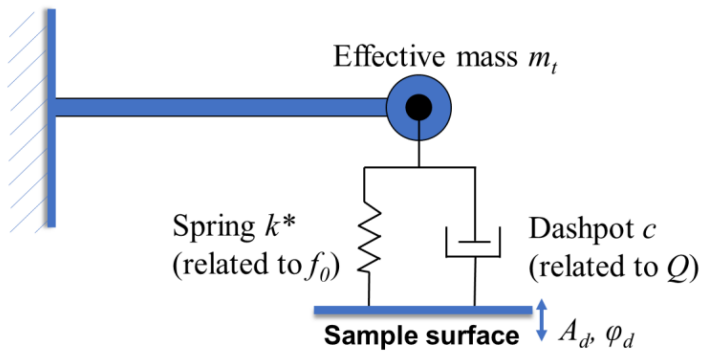

**Figure 1.** A schematic diagram showing the tip-sample oscillating system: the forces between the tip and the sample surface can be represented by the spring $k^*$, and the damping can by represented by the dashpot $c$. $m_t$ is the effective mass of the tip. $k^*$ and $c$ are related to the contact resonance frequency ($f_0$) and quality factor ($Q$), respectively. $A_d$ and $\phi_d$ are the driving amplitude and phase of the sample surface, respectively.

In the pulsed DC mode, the piezoresponse are measured when the switching DC bias is on (on-field) or off (off-field).[37] In the on-field, the applied DC voltage induces the piezoelectric motion of the domain and domain walls, then ferroelectric materials keep this stable status in the following off-field. The off-field signal is usually considered as the clear response of the tip-sample interaction without the influences from strong DC field-induced tip-sample electrostatic interaction. Generally-speaking, PFS measurements can obtain the changes of amplitude ($A$) and the phase angle ($\phi$) as functions of the DC bias.[38] The piezoresponse ($PR$), as a function of DC bias, can be then calculated by the following equation:[38]

$$PR = A \cdot \cos(\phi) \tag{6}$$

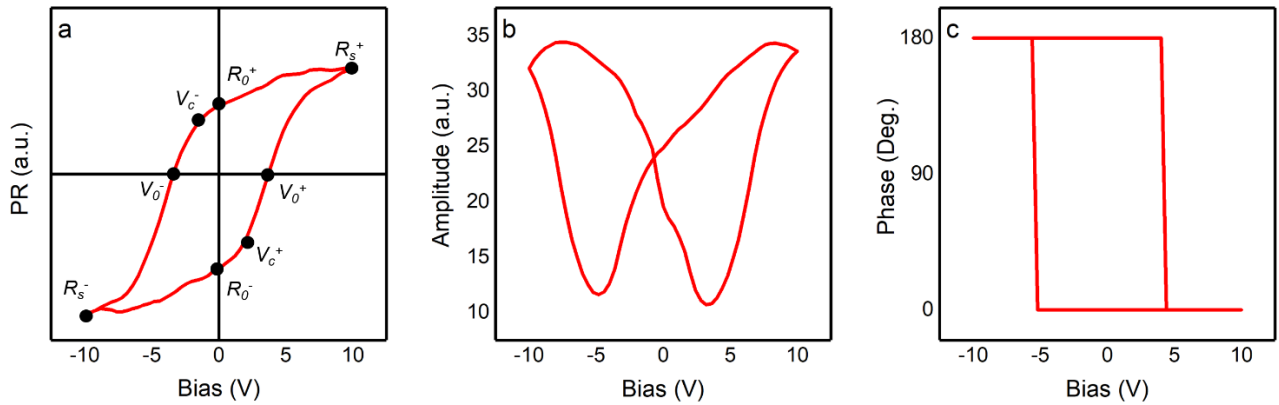

**Figure 2.** A schematic of (a) calculated *PR* loop; (b) "butterfly-shape" amplitude loop; and (c) phase loop, all for ferroelectric material. In the *PR* loop showed in (a), $V_0^+$ and $V_0^-$ are forward coercive and reverse coercive biases, respectively, at which the *PR* responses are equal to zero. $V_c^+$ and $V_c^-$ are nucleation biases, representing the initiation of the domain switching processes in the ferroelectric materials, $R_0^+$ and $R_0^-$ are forward and reverse saturation piezoelectric responses. $R_s^+$ and $R_s^-$ are remnant piezoelectric responses.

Figure 2 shows a schematic of "butterfly-shaped" amplitude loop, as well as the phase transition loop and calculated *PR* loop from the PFS measurements in ferroelectric materials. Due to the nonlinear piezoelectric responses, *PR* curve forms a closed hysteresis loop under the cyclic DC voltage sweeping, which is regarded as a general electromechanical response from ferroelectric materials.[39] The shape of an electromechanical hysteresis loop depends on the properties of the material and the experimental conditions.[40] In other words, all these parameters are associated with the true physical properties of the materials. Therefore, yielding a hysteresis loop in PFS measurement in the off-field is generally a well-recognized evidence for ferroelectricity on the range from nanoscale to macroscale.[40–42]

**Experimental observation.** When conducting PFS experiments, it is noted that the hysteresis loops can be observed in both relaxor PZN-PT and banana peel sample. PZN-PT is a typical relaxor ferroelectric material[43] and banana peel is a non-ferroelectric material.[5] By carefully observing and comparing all of the responses from the PFS measurements of the PZN-PT and banana peel samples, it is found that the signals of $f_0$ spectrum show distinctly different features in the two materials during the DC bias cycle (Figure 3). However, the amplitude, phase and *PR* responses from both materials show similar shape. . It is therefore ambiguous to differentiate these two materials by using the hysteresis loops only, but $f_0$ can be a distinct parameter to prove true ferroelectric response. After the preliminary observation, more materials are measured with the PFS techniques and then based on the experimental findings, we hence propose a simple but effective method based on $f_0$ from the PFS measurement to distinguish the responses from the ferroelectric and non-ferroelectric materials.

**Data analysis.** In order to investigate the relationships between *PR* and $f_0$ in ferroelectric and non-ferroelectric materials, we first re-plot the data as *PR* versus $f_0$ plots. In this plot, the x-axis is *PR* responses (calculated from Eq.(6) based on experimentally obtained PFS amplitude and phase angle) and the y-axis is the $f_0$ values during the same PFS measurements. The local off-field hysteresis loops and amplitude loops of the ferroelectric materials can be seen in Figures 4(a), (d), (g) and (j). The deformations of the samples can be reflected by the normalized signals of amplitude (*A*). The shapes of the off-field piezoresponse loops and the amplitude loops have slight differences among these samples. The relationships between off-field *PR* and $f_0$ of these materials are shown in Figures 4(b), (e), (h) and (k). Remarkably, all ferroelectric materials show two sharp peaks in the *PR-$f_0$* curves at the position where *PR* nearly equals to zero. Those peaks are marked by red dots in the figures, the corresponding positions in the hysteresis loops and amplitude loops are also marked (by black dots on hysteresis loops and by blue square dots on the amplitude loops) as well. The $f_0$ peaks in the on-field *PR-$f_0$* curves are very sharper and narrower, which can be seen in Figures 4(c), (f), (i) and (l). The positions of the on-field $f_0$ peaks are even closer to the zero piezoresponse compared with that of the off-field signals.

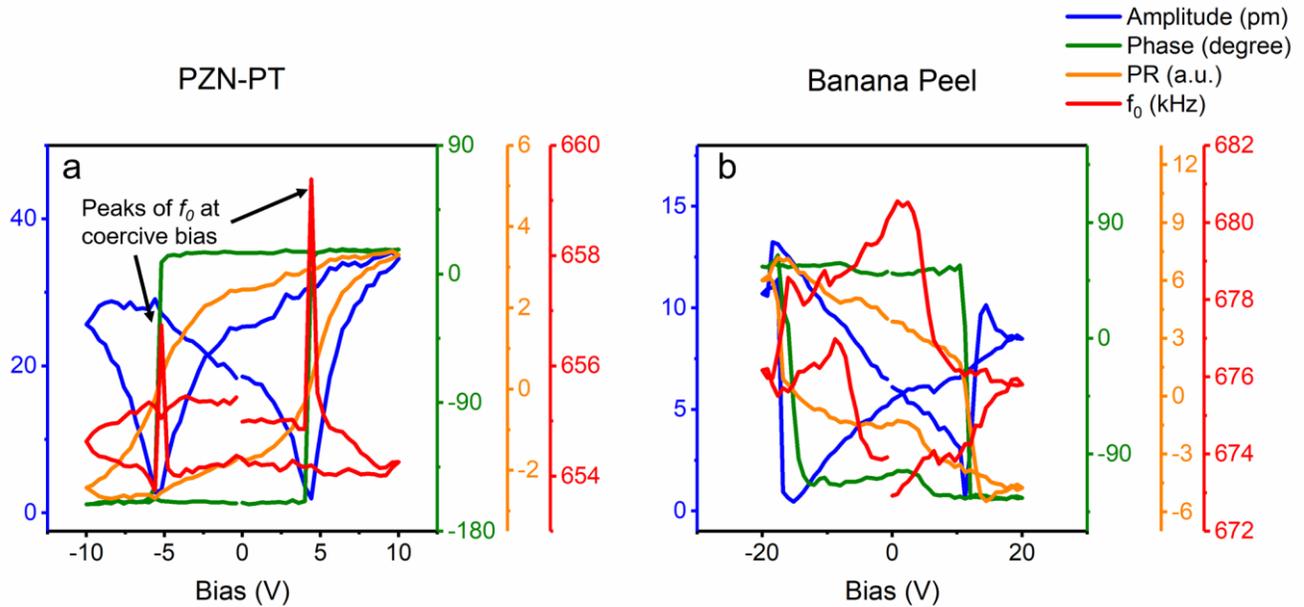

**Figure 3.** Comparison of the PFS measurements between (a) ferroelectric material (PZN-PT) and (b) non-ferroelectric material (banana peel). The PFS measurements include the amplitude loop (blue line and blue scale), phase loop (green line and green scale), the calculated piezoresponse loop (orange line and orange scale), and the changes of $f_0$ during the PFS measurements (red line and scale). Noted the two sharp peaks in the contact frequency curve (indicated by arrows) for PZN-PT whereas no clear peaks or patterns for $f_0$ banana peel.

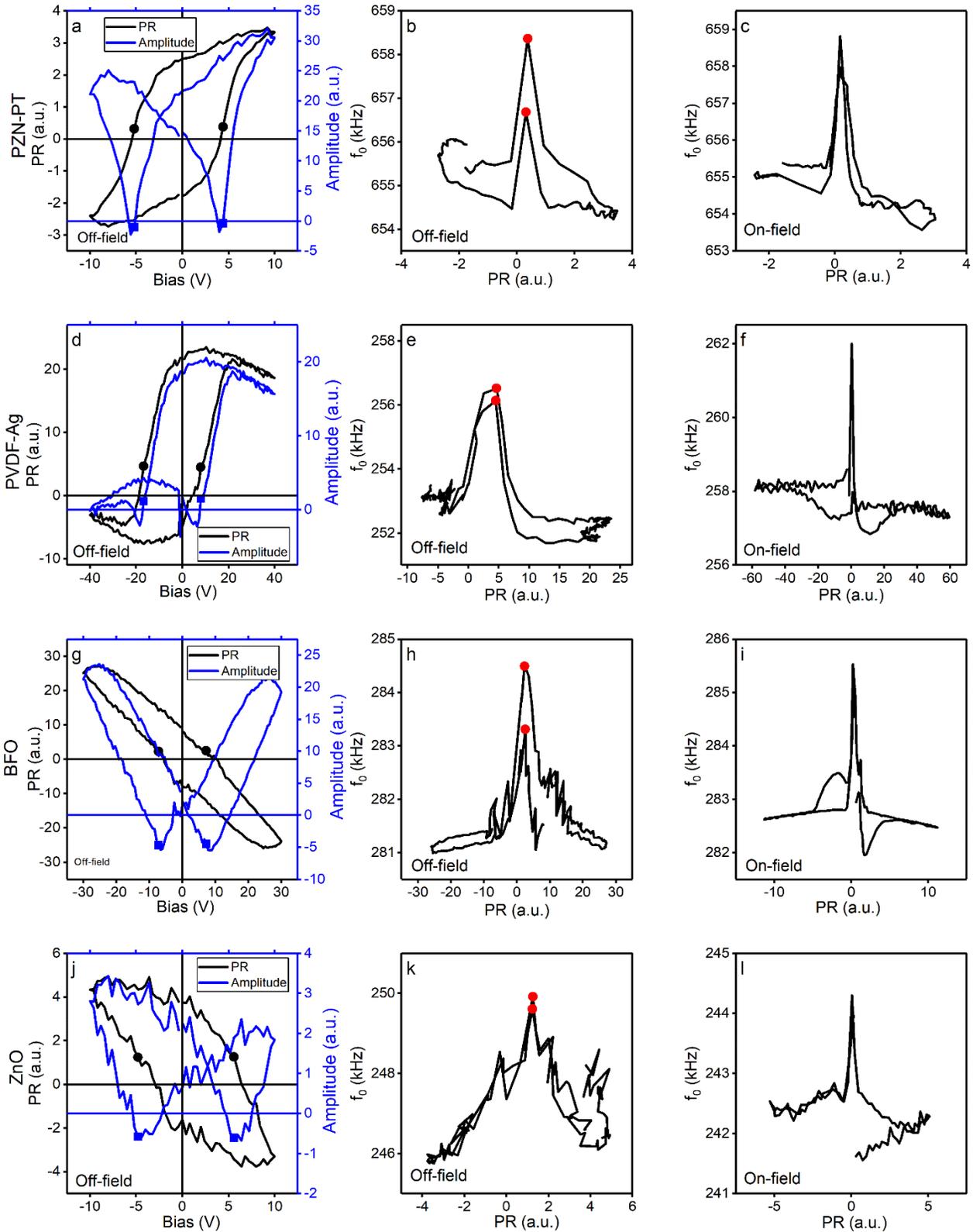

**Figure 4.** PFS amplitude loop (measured at off-field) and calculated hysteresis loop for ferroelectric materials: (a) PZN-PT, (d) PVDF-Ag, (g) BFO and (j) ZnO. Contact resonance frequency as function of calculated piezoresponse (*PR-f$_0$*) for (b) PZN-PT, (e) PVDF-Ag, (h) BFO and (k) ZnO at off-field and (c), (f), (i) and (l) obtained for the same materials at on-field. The red dots in (b), (e), (h) and (k) show the peak positions in the *PR-f₀* loop. The points marked by black or blue color in (a), (c), (e) and (g) show the corresponding piezoresponse and amplitude where the contact resonance frequencies reach to the peak values, respectively.

It is illustrated that, in general, the $f_0$ remains at a constant with a small value, however, when the applied DC voltage reaches to the coercive bias, $f_0$ values for ferroelectric materials, namely PZN-PT, PVDF-Ag, BFO and ZnO, jump to a notably high value suddenly and when the applied DC voltage is higher than the coercive bias, $f_0$ reverts to the values as before. It is also noticed that the slope of *PR* curve reaches to the maximum at the same time. In the saturated response region, $f_0$ is almost a constant. This pattern of the change repeats when *PR* reverses due to the changes of the direction in the applied electric field. Comparing the off-field plots and on-field plots, the electrostatic force affects not only the *PR* signals, but also $f_0$ signals. However, both the off-field and on-field plots show obvious common patterns of $f_0$ in all four ferroelectric materials during the PFS measurement.

It is known that some non-ferroelectric materials also demonstrate ferroelectric-like hysteresis loops and amplitude loops during the PFS measurements, such as glass and banana peel, which can be seen in Figures 5(a) and (d). The minimums of the amplitude are also around zero in glass and banana peel. However, both the off-field and on-field *PR-f₀* curves show significant differences between the ferroelectric materials and the non-ferroelectric materials. Most importantly, no $f_0$ peaks are observed in the PFS measurements of the glass and banana peel samples at the positions around coercive bias, for both off-field and on-field signals. In Figures 5(b), (c), (e) and (f), with the biases forwarding and reversing, the off-field and on-field *PR-f₀* curves of glass and banana peel are completely different with those of the ferroelectric materials showed in Figure 4. Each of the non-ferroelectric material has random $f_0$ signals. Hence it is suggested that their hysteresis loops are not ascribed to the polarization reversing or switching. The factors which contribute to the hysteresis-like loops in glass and banana peel are most likely not the same, presumably by the dissimilarity of *PR-f₀* curves. Actually, the PFS signals of non-ferroelectric materials are strongly affected by the experiment conditions and it is hardly to obtain the consistent results in every repetitive experiment.

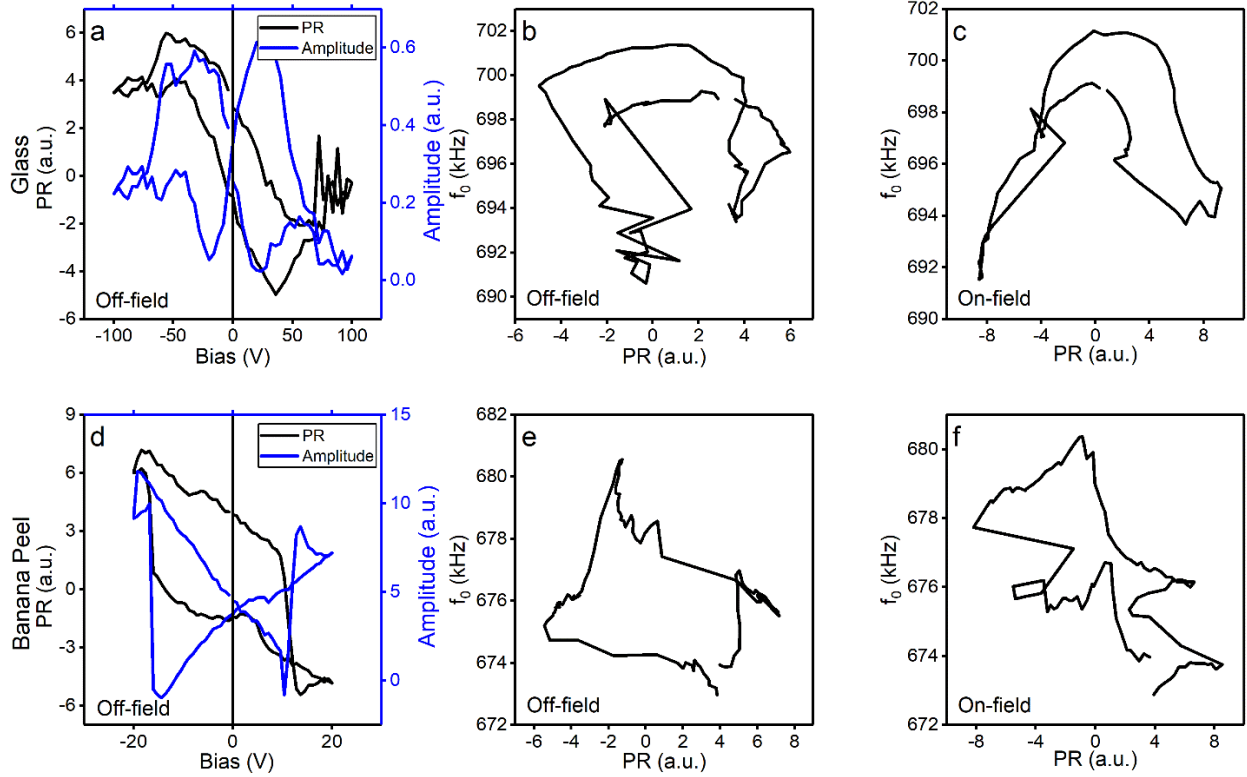

**Figure 5.** PFS amplitude loop (measured at off-field) and calculated hysteresis-like loops for non-ferroelectric materials with PFS measured amplitude and phase loops: (a) glass and (d) banana peel. Contact Resonance Frequencies as function of calculated piezoresponse (*PR-f₀*) loop for (b) glass and (e) banana peel measured at off-field and (c) and (f) for the same materials measured at on-field, respectively. Note there are no contact resonance frequency peaks and any regular patterns for the curves in those materials.

Furthermore, we also conducted the PFS measurements on two other non-ferroelectric materials (bulk PMMA and Si), PFS measurements cannot get any hysteresis-like loops in these two materials. As expected, the $PR$-$f_0$ curves are highly random and no peaks can be observed. Their behaviours during the PFS measurements are also clearly differentiated from that of the ferroelectric materials. As no hysteresis loops can be observed in those materials, the curves are hence not showed here.

In order to investigate the endurance of $f_0$, we also conduct multicycles PFS measurements for both ferroelectric and non-ferroelectric materials. Ten (10) cycles of PFS measurements have been conducted for both materials. The original measured $f_0$ signals are constituted with a lot of non-linear details, for example the peaks at coercive bias. We therefore use wavelets analysis (using Matlab) to remove these details and only focus on the main trend of $f_0$ signals as functions of PFS cycles. The sixth order approximation signals after Wavelet Daubechies (db4) transform are shown in Figure 6 for both ferroelectric and non-ferroelectric materials. Daubechies Wavelets, known as "compact support orthogonal wavelets", can decompose data into approximations and details without gap or overlap, in which is usually used to detect or filter the nonlinear or instantaneous response signal processing.[44] To obtain the clear trend of each sample and compare them, the signals have been normalized by the initial value of the time sequence during the PFS measurement. In all the cases, $f_0$ signals are unstable in the first cycle, but tend to be stable after 2 or 3 cycles. Comparing three repeated measurements on PZN-PT samples, the shifts of $f_0$ in the early cycles are not identical. It may be related to the pre-set drive frequency of the PFS measurement. For glass and banana peel in the PFS measurements, the patterns of the $f_0$ under repetitive cyclic field are more similar to the non-ferroelectric material, PMMA, which do not show a hysteresis loop of piezoresponse in the ten cycles.

**Discussion.** Through the experiments and analysis of the above materials, it is obvious that, for ferroelectric materials, $f_0$ values at both off-field and on-field change to a peak value when the applied bias reaches to the coercive bias, at this moment, the piezoresponse is equal to zero except the sample of doped ZnO. ZnO is not a traditional ferroelectric material. During the PFS measurements, not all of the points can obtain the hysteresis loop. However, at the locations that can detect the ferroelectric-like responses, $f_0$ shows similar behaviour as those in the traditional ferroelectric materials. The corresponding biases deviate from the coercive biases slightly because the *PR* values are larger than zero at the positions of the two peaks, which may be related to the atomic structure of the doped ZnO. At coercive bias, the amplitude-electric field loop also reaches to the minimum value that is slightly less than zero. Generally-speaking, during the PFS measurements, the $f_0$ increases to a peak value suddenly when the deformation reverses to zero in one testing cycle. When the applied bias is higher than the coercive bias, the $f_0$ reduces back to the smaller value immediately. Such patterns of the $f_0$ versus bias appear in all of four ferroelectric materials studied in this work for both off-field and on-field measurements, although the magnitudes of the $f_0$ are different for different materials. The on-field $f_0$ is affected by the electrostatic force, but the peak appears enhanced. On the other hand, non-ferroelectric materials, such as glass and banana peel, have more straightforward phase flipping in on-field, but like their off-field signals, on-field signals do not show any regular patterns in terms of $f_0$. For all experiments, the magnitudes of $f_0$ are strongly affected by the stiffness of the tips and samples as well as the pre-set values of the PFS measurements.

The bias-induced deformation in piezo-/ferro-electric material is caused by the converse piezoelectric effect of the lattice as well as the switching and movement of domain walls.[45] For ideal ferroelectric materials, the lowest value of deformation or strain in a cycle should be the strain with the most negative value.[46] However, for the ferroelectric materials studied here, the intersection point of the "butterfly" is above zero due to the effects of the residual strain.[47] The magnitude of the maximum positive strain is much higher than the magnitude of the maximum negative strain which is close to zero value. For some non-ferroelectric materials, such as glass and banana peel, the bias-amplitude loops are also "butterfly" shape, which is similar to that of the ferroelectric materials. However, these materials do not have domain walls and bias-induced lattice deformation, and certainly, the deformation or strain is not induced by piezoelectric effects, hence the amplitude changes during the PFS measurement are induced by the factors other than ferroelectric behavior, such as surface charge and Vegard effects.[5,15,23]

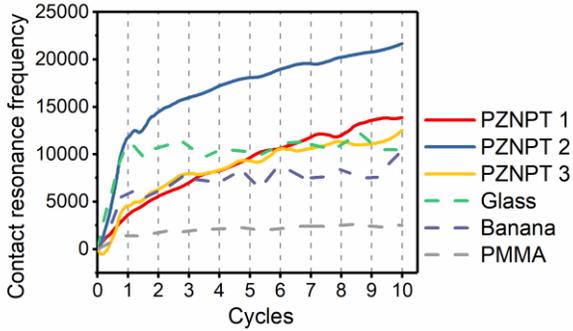

**Figure 6.** Analysis of the trend of the contact resonance frequency($f_0$) curves for both ferroelectric and non-ferroelectric materials by using wavelet transformation: 10 cycles of PFS measured $f_0$ signals (measured at off-field) after 6th approximation of the wavelets transform. The solid lines represent three individual tests on PZN-PT (ferroelectric material) samples. The dash lines represent the test data on non-ferroelectric materials of glass, banana and PMMA, respectively. Note the trend of the contact resonance curves of the PZN-PT show increasing continuously with the testing cycles, whereas for non-ferroelectric materials, after the initial cycle, the trend of $f_0$ curves become independent of the testing cycles. The wavelet analysis is performed by using Matlab (R2016b).

From the analysis of the responses of different materials during the PFS measurements, the signal of $f_0$, as a parameter which can represent the mechanical properties of the sample, show remarkable differences under the electric field between the ferroelectric and non-ferroelectric materials. These observations signify there is a notable increase of the contact stiffness during the polarization switching processes in ferroelectric materials, which does not occur in non-ferroelectric materials. Referring to the physical meanings of the changes of the $f_0$ values, it is believed that this sharp increase is caused by an instantaneous increase of Young's modulus of the sample at the moment when new domain is nucleated. Eq.(2) illustrates that the tip-sample interaction force is mainly contributed by Hertzian contact force. Eqs.(2) and (3) also show that the interaction force is affected by Young's modulus and Poisson's ratio of the tip and the sample. Therefore, the increase of Young's modulus of the sample can lead to an increase of the contact stiffness and hence the contact resonance frequency. In addition, Eq.(5) shows that, if the peaks of the $f_0$ are not so sharp or high in ferroelectric materials, the reason may be related to the ratio of the contact stiffness to cantilever stiffness.

In the earlier studies of the constitutive model for ferroelectric materials, it was found that the work-hardening effects could not be neglected.[48] A small hardening rate existed during the bias cycle processes. In this study, it is found that the contact resonance frequencies of PZN-PT increase constantly under the repetitive cyclic field (Figure 6), which may be related to the hardening effects during the polarization switching. On the other hand, this hardening effect does not exist in non-ferroelectric materials. Based on the differences between the ferroelectric materials and non-ferroelectric materials during the multiple cycles of PFS measurements, we believe that the hardening processes of the ferroelectric materials induced by polarization switching can also be reflected on the signals of the $f_0$ from the PFS measurements. Furthermore, the constitutive model also assumed that the hardening effects by 90° domain switching differ from that by 180° domain switching.[48] It can be extrapolated that the slope of the $f_0$ increasing depends on the angle of the local domain. Combined with the analysis of the one-cycle results and ten-cycle results, it is believed that the hardening of ferroelectric materials is a non-linear process. A significant hardening occurs at the coercive bias and then the Young's modulus of the sample almost totally revert to the former status but with a small amount of increase in the hardening effect which will not disappear and can be accumulated during the subsequent cycles. Owing to the very small strain ($\varepsilon$) at the coercive bias, though the Young's modulus ($E$) increases to a notably high value, the stress ($\sigma$) remains small. It ensures less destruction during the non-linear hardening processes because the internal stress is not extremely large. On the other hand, in non-ferroelectric materials, the $f_0$ changes randomly with the poling voltage, which indicates no polarization switching processes even PFS measurements can get a similar hysteresis loop in those materials.

Due to the obvious differences in the changes of the off-field and on-field $f_0$ values between the ferroelectric and non-ferroelectric materials, it is therefore possible to differentiate responses from these two types of materials by observing the distinct $f_0$ peaks in the $PR$-$f_0$ curves. Apparently, this is a simple yet effective method to differentiate the two types of materials' responses even though the PFS measurements give the similar hysteresis loops in these materials.

SUMMARY AND CONCLUSION REMARKS

In summary, this study develops a new, simple yet effective method for differentiating between the similar responses from the ferroelectric and non-ferroelectric materials during the PFS measurements. Different from the traditional method to distinguish ferroelectricity by local hysteresis loop that is only constituted by the amplitude and the phase angle, this new method pays attention to the changes of the contact resonance frequency during the PFS measurements, in which provides a new information about the change of mechanical properties during the domain switching processes with remarkably differences between the ferroelectric and non-ferroelectric materials. In this study, two groups of materials, including four ferroelectric materials (PZN-PT, PVDF-Ag, BFO, doped ZnO), two non-ferroelectric materials (glass and banana peel) showed hysteresis loops, and an additional group of two non-ferroelectric materials (PMMA and Si) that have no hysteresis loops, are studied by PFS experiments and the results are analyzed using this new method to support the conclusions. The results prove that the $f_0$-based method is very robust and effective. All four ferroelectric materials show the similar regular patterns in term of the changes of the contact resonance frequency during the PFS measurements, whereas the four non-ferroelectric materials do not show any regular patterns regardless whether they can obtain the hysteresis loop during the PFS measurements.

In other modes of SPM techniques, such as contact resonance frequency AFM (CR-FM), the contact resonance frequency is usually used to measure elasticity and viscoelasticity.[49] Although the excitation method in PFS is different from that in the CR-FM technique, the contact resonance frequency still depends on the contact stiffness. From analysis of experimental results, the differences of the instantaneous response of $f_0$ between the ferroelectric materials and non-ferroelectric materials are obvious, and this is easy to distinguish the differences between the ferroelectric and non-ferroelectric materials during the nanoscale PFS measurements, which illustrates a non-linear hardening processes during the domain switching. Combined with the theory of the contact resonance frequency in the tip-sample oscillating system, it is suggested that Young's modulus of the sample changes instantaneously at the coercive bias. Our new method can verify the ferroelectric behaviors by examining the contact resonance frequency peaks appearing in the $PR$-$f_0$ curves. This method not only provides a new and simple way to differentiate ferroelectric and non-ferroelectric materials, but also presents a new direction to characterize the ferroelectric responses. We also speculate that it is possible to use the changes of the contact resonance frequency during the PFS measurements to characterize the domain switching dynamics in the ferroelectric materials.

MATERIALS AND EXPERIMENTS

In this study, three groups of eight materials were tested, including four ferroelectric materials, two non-ferroelectric materials with PFS measured hysteresis loops, and two non-ferroelectric materials without any hysteresis loops can be measured from the PFS experiments. Those materials are selected to investigate $f_0$ ferroelectric and non-ferroelectric materials during the PFS measurements. The ferroelectric materials include $Pb(Zn_{1/3}Nb_{2/3})O_3$–9%$PbTiO_3$ (PZN–PT) single crystals, hybrid polymeric–metallic (PVDF–Ag) composite, $BiFeO_3$ (BFO) thin film and Cu-doped ZnO film. We have reported the ferroelectric characteristics of PZN-PT[43] and PVDF-Ag[50] in the previous studies, thus those materials are considered as typical examples of the ferroelectric materials. BFO is another classic ferroelectric material which shows the ferroelectric properties in room temperature and there are three types of domains with different angles, 71°, 109° and 180°.[51] It is different from that of the PZN-PT and PVDF-Ag, in which the latter two materials show only 180º domains.[6,52–55] Cu-doped ZnO film has been observed with ferroelectric-like behaviors in our previous studies, which with small grain size (around 25 nm) showed an effective polarization switching under the applied electric field.[6,52–55] The non-ferroelectric samples are glass and banana peel, which were reported to show the hysteresis loop during the PFS measurements.[5,15] Bulk Si and Poly(methyl methacrylate) (PMMA) are non-ferroelectric materials that hardly observe any hysteresis loop during the PFS measurements.

The PFS measurements were conducted in all eight materials. The amplitude, phase and contact resonance frequency of the eight materials were measured using a commercial SPM system (MFP-3D, Oxford Instruments, CA, USA). The PFS measurements were conducted under the Dual-AC Resonance Tracking (DART) mode. The PFS amplitude and phase loops were obtained from the off-field and on-field respectively. The off-field and on-field *PR* loops were calculated from the corresponding measured amplitude and phase loops according to Eq.(6). The corresponding on-field and off-field contact resonance frequencies were also obtained at the same time from the PFS measurements.

## AUTHOR INFORMATION

### Corresponding Author

*Prof. Dr. Kaiyang Zeng, E-mail: mpezk@nus.edu.sg, Tel: (+65) 6516 6627, Fax (+65) 6779 1459.

### Author Contributions

The manuscript was written through contributions of all authors. YL performed all of the numerical analysis and wrote this manuscript. Other authors were contributed to the large amount of PFM and PFS experiments on various materials and discussions about the results and manuscript. All authors have given approval to the final version of the manuscript.

### Funding Sources

This work is supported by Ministry of Education (Singapore) through National University of Singapore (NUS) under the Academic Research Fund (ARF), R-265-000-596-112.

# Notes

The authors declare no competing financial interest.

ACKNOWLEDGMENT

The authors would like to thank the financial support by Ministry of Education, Singapore, through National University of Singapore (NUS) under the Academic Research Fund (ARF) of grant number R-265-000-596-112. The authors (YL, YS, ZTW, BXY) also thanks the post-graduate scholarship provide by National University of Singapore.

ABBREVIATIONS

PFS, Piezoresponse Force Spectroscopy; PFM, Piezoresponse Force Microscopy; P-E, polarization-electric; SS-PFM, switching spectroscopy Piezoresponse Force Microscopy; SHG, second harmonic generation; cKPFM, contact Kelvin Probe Force Microscopy; DHO, damping harmonic oscillator; SPM, Scanning Probe Microscopy; DART, dual AC resonance tracking; BE, band exciation; SNR, signal to noise ratio; AFAM, Atomic Force Acoustic Microscopy; AFM, Atomic Force Microscopy; PR, piezoresponse; PZN-PT, $Pb(Zn_{1/3}Nb_{2/3})O_3-9\%PbTiO_3$; CR-FM, contact resonance frequency Atomic Force Microscopy; PVDF-Ag, hybrid polymeric-metallic; BFO, $BiFeO_3$; PMMA, Poly(methyl methacrylate).